\documentclass[prd,amsfonts,onecolumn,superscriptaddress,aps,nofootinbib,11pt]{revtex4-1}

\usepackage{ragged2e}
\usepackage[utf8]{inputenc} 
\usepackage[T1]{fontenc}    
\usepackage{hyperref}       
\usepackage{url}            
\usepackage{booktabs}       
\usepackage{amsfonts}       
\usepackage{amsmath,amssymb}
\usepackage{nicefrac}       
\usepackage{microtype}      
\usepackage{lipsum}
\usepackage{graphicx}
\graphicspath{ {./images/} }
\usepackage{lipsum}  
\usepackage{caption}
\usepackage{subcaption}
\usepackage{graphicx}
\usepackage{dcolumn}
\usepackage{bm}
\usepackage{tikz}
\usepackage[compat=1.1.0]{tikz-feynman}
\pgfkeys{/tikzfeynman/warn luatex=false}
\usepackage{slashed}
\usepackage{float}

\begin{document}

\title{Very Special Relativity: Cherenkov Effect and an Analogy with Minkowski's Electrodynamics of Continuous Media
}

\author{I. H. Brevik}
\email{iver.h.brevik@ntnu.no}
\affiliation{Department of Energy and Process Engineering, Norwegian University of Science and Technology, N-7491 Trondheim, Norway.}

\author{M. M. Chaichian}
\email{masud.chaichian@helsinki.fi}
\affiliation{Department of Physics and Helsinki Institute of Physics, University of Helsinki, P.O. Box 64, 00014 Helsinki, Finland.}

\author{B. A. Couto e Silva}
\email{brunoaces@ufmg.br}
\affiliation{Department of Physics and Helsinki Institute of Physics, University of Helsinki, P.O. Box 64, 00014 Helsinki, Finland.}
\affiliation{Departamento de F\'isica, UFMG, Belo Horizonte, MG 31270-901, Brazil.}

\author{B. L. S\'anchez-Vega}
\email{bruce@fisica.ufmg.br}
\affiliation{Departamento de F\'isica, UFMG, Belo Horizonte, MG 31270-901, Brazil.}

\date{\today}

\begin{abstract}
In this work, we explore the implications of the Cohen and Glashow Very Special Relativity (VSR) theory, a framework that introduces Lorentz invariance violation through the presence of a preferred direction. Our analysis focuses on the impact of VSR on the Cherenkov angle, revealing modifications to the dispersion relation of particles, particularly the photon and the electron, which acquire an effective inertial mass. This modification also implies a deviation in the speed of light, which can be constrained through precise experimental measurements. Using data from the RICH system of the LHCb experiment, we take advantage of its capability to reconstruct Cherenkov angles within the momentum range of the particles of 2.6–100 GeV/c. These measurements, combined with the most stringent laboratory tests of the isotropy of the speed of light ($\Delta c / c \sim 10^{-17}$), allow us to impose new upper bounds on the parameter $\Omega$, which quantifies a deviation from the standard Special Relativity. Furthermore, we establish an analogy between VSR and Minkowski's electrodynamics in a dielectric medium for particles with very high velocity, offering a physically intuitive interpretation of the parameter $\Omega$.
\end{abstract}

\maketitle

\tableofcontents


\newpage
\section{Introduction}

The search for a theory that unifies general relativity and quantum mechanics has led physicists to explore various candidates for Planck-scale physics \cite{ProceedingsL}. Among these candidates, Lorentz and CPT (Charge, Parity, Time reversal) violations have gained significant interest. These violations, if observed, could provide critical insights into the nature of spacetime at the smallest scales and offer clues about the underlying structure of the Universe.

Lorentz Invariant Violating (LIV) theories, were largely driven by the discovery of parity violation in weak interactions \cite{LeeYang1956,Wu1957}. If parity, once believed to be an exact symmetry, is broken, it raises the possibility that other fundamental symmetries, like the Lorentz symmetry, may also be violated under certain conditions.

One of the possibilities to violate the Lorentz symmetry is known as noncommutative quantum field theories \cite{Seiberg1999}. In this framework, the structure of spacetime is modified such that the spacetime coordinates no longer commute, but instead follow a relation like $[x_\mu,x_\nu]=i\theta_{\mu \nu}$, where $\theta_{\mu \nu}$ is a constant matrix that determines the scale of non-commutativity. This novel structure changes the fundamental nature of spacetime at small scales, leading to a modification of how fields and particles interact. In particular, the non-commutative structure breaks the usual Lorentz symmetry since the matrix $\theta_{\mu \nu}$ introduces a preferred direction or scale in spacetime, thus violating Lorentz invariance. There are several other approaches to introduce a mechanism to break the Lorentz invariance (see, e.g. \cite{klink} and references therein).

Taking a different approach, Coleman and Glashow \cite{Coleman1999} proposed that the Lorentz symmetry could be violated at extremely high energies, discussing the implications of Lorentz symmetry violation and proposing methods for testing their hypothesis using Ultra High-Energy Cosmic Rays.

A LIV theory that has been receiving significant attention is the Cohen and Glashow Very Special Relativity (VSR) theory \cite{VSRGC}, which is constructed from symmetries defined within subgroups of the Poincaré group. This framework encompasses not only spacetime translations but also at least a two-parameter proper subgroup of the Lorentz group. We notice that a  realization of the Cohen-Glashow Very Special Relativity has been given within the noncommutative spacetime \cite{Tureanu}.

VSR modifies the traditional understanding by restricting the full Lorentz symmetry into one of their subgroups, however, if one introduces the discrete symmetries $T,P$ or $CP$ to those subgroups, it will recover the full Lorentz group. It is interesting to observe that if CP were an exact symmetry of Nature, it would imply that inserting a CP transformation (which swaps particles with their antiparticles and reflects spatial coordinates) on the symmetry group of Very Special Relativity would extend it to the full Lorentz group. In other words, under exact CP symmetry, VSR would be indistinguishable from the standard Lorentz transformations of Special Relativity (SR) i.e. all effects predicted by SR would hold universally, with no deviations. However, CP symmetry is not exact \cite{CPPRL,CPPRL2}, meaning that the transformation between VSR and the full Lorentz group is incomplete, leaving room for subtle deviations from SR. Since the effects that violate the CP symmetry are inherently small (only appearing in specific interactions at the quantum level), we expect that any deviations from SR, that VSR predicts, will also be small, proportional to a parameter that we will call $\Omega.$

This modification aims to address unresolved issues in particle physics and cosmology while preserving the fundamental experimental results that support Special Relativity. By choosing a subset of the Lorentz group, the authors discuss the novel physics that can emerge in the VSR framework \cite{CGnm}. One of them concerning about the mass of the neutrinos that are massless in the Standard Model, however, experiments have shown that neutrinos oscillate between different flavors, implying that they have mass \cite{Adamson2011,Takeuchi2012}. The VSR scenario could influence neutrino oscillations and proposes that neutrino masses and mixing angles could arise naturally from the symmetries of the theory, providing a novel explanation for these phenomena without invoking new particles or forces. This approach contrasts with the need for additional particles or forces to explain neutrino oscillations in the context of SR, as presented in \cite{Gouvea} and references therein.

The most significant difference between VSR and SR is the symmetry groups they employ. Special Relativity relies on the Lorentz group, encompassing rotations and boosts in all directions. In contrast, VSR proposes smaller subgroups like $SIM(2)$ or $HOM(2)$, which maintain some but not all Lorentz invariance properties. The $SIM(2)$ group, for example, includes rotations and boosts in a specific plane, preserving the constancy of the speed of light in that plane but allowing for potential deviations in other directions. That could, in principle, lead to observable effects that differ from those predicted by SR, such as, the speed of light varying slightly depending on the direction of propagation, and this variation could be measured using highly sensitive instruments, providing a potential test for the validity of VSR. 

An interesting feature of this proposal is the existence of a null-vector, which from it, one can construct nonlocal operators that violate Lorentz invariance while respecting $SIM(2)$ or $HOM(2)$ dot products of this vector with kinematic variables. Thus, results may depend on the direction relative to the VSR preferred direction. Despite the presence of a preferred direction, the theory still aims to uphold the principle of relativity, which states that the laws of physics should be the same in all inertial frames of reference. This is achieved in VSR by ensuring that while there is a preferred direction, the transformations between different inertial frames still preserve the form of the physical laws. 

It is important to note that, since we are working with a subgroup of the Lorentz group, the representations of the theory differs. In Very Special Relativity, the representation theory of \( SIM(2) \) leads to one-dimensional representations for massive particles. In these representations, spin states along a preferred axis can have different masses. These spin-dependent mass terms introduce a coupling that is tightly constrained by experimental data, raising concerns about a fine-tuning problem, as discussed in \cite{repspin}. However, we will not explore this issue further, as the propagators derived below are consistent with the representation structure of \( SIM(2) \), that can generate physical particle states. 

Our work is organized as the following: in Section \ref{section 1}, we investigate and evaluate the implications of the Very Special Relativity theory by analyzing the effects of the $\Omega$ parameter on the Cherenkov angle. Specifically, we will calculate how modifications introduced by the VSR framework alters this radiation, following the methodology outlined in \cite{Cox}. By deriving an expression for the Cherenkov angle within the VSR context, we are able to set bounds on the value of $\Omega$.

In Section \ref{section 2}, we turn our attention to the reduction of the speed of light as predicted by the VSR framework. Building on this premise, we will estimate upper bounds on the $\Omega$ parameter by examining experiments on the constancy of the speed of light. This analysis  provides experimentally testable predictions and the derived bounds on $\Omega$ offer critical insights into the parameter's constraints and its potential effects on detecting Lorentz invariance violation signatures from VSR.

Finally, in Section \ref{section 3}, we present an analogy between the Very Special Relativity and Minkowski’s electrodynamics in dielectric media. Our findings show that, for a medium moving with a high velocity, the photon dispersion relation in Minkowski electrodynamics is altered in a manner analogous to the VSR scenario. This provides a clearer physical interpretation of VSR, as will be presented.

\section{The Cherenkov Effect in the Very Special Relativity framework}\label{section 1}

Consider a VSR Lagrangian invariant under $SIM(2)$ transformations \cite{CGnm,Cheon2009}, given by
\begin{equation}
    -\frac{1}{4}\Tilde{F}_{\mu \nu}\Tilde{F}^{\mu \nu}+\frac{1}{2 \xi}(\partial^\mu A_\mu)^2 + J^\mu A_\mu,
\end{equation}
where $\Tilde{F}_{\mu \nu} \equiv \Tilde{\partial}_\mu A_\nu - \Tilde{\partial}_\nu A_\mu$ and the wiggle derivative is defined as $\Tilde{\partial}_\mu \equiv \partial_\mu - \frac{1}{2}\frac{\Omega^2}{s \cdot \partial}s_\mu,$ where the null four-vector is written as $s=(1,0,0,1),$ with the preferred direction chosen to be the z-axis. A similar structure was also described and constructed in \cite{Bauer2001}, where the nonlocal terms in both the heavy quark effective theory and in Very Special Relativity are similar in that they both introduce modifications to local interactions through terms that involve constraints over specific directions.

In Very Special Relativity, the non-local operator $\frac{1}{s \cdot \partial}$ vanishes for large momentum due to the growth of $s \cdot \partial$ in momentum space. This suppression exhibits small effects at high energy scales, where shorter wavelengths dominate, showing that the VSR modifications are negligible and in accordance to the standard local field theory in the ultraviolet (UV) regime. However, at low momentum (infrared, or IR, scales), $\frac{1}{s \cdot \partial}$ becomes significant in momentum space, potentially leading to IR divergences. These divergences are a feature of the non-local structure in VSR and play a critical role in the infrared behavior and phenomenological implications of the theory.

The dispersion relation for the photon deviates from the standard relativistic framework. This can be seen from the propagator for the gauge sector that, when choosing the Feynman gauge $(\xi=1),$ is given by 
\begin{equation}
\begin{aligned}
   &\Delta_{\mu \nu}=-\frac{i}{k^2-\Omega^2}\left[\eta_{\mu \nu}+\frac{\Omega^2}{(s \cdot k)^2} s_\mu s_\nu -\frac{\Omega^2}{k^2 (s \cdot k)}\left(k_\mu s_\nu+k_\nu s_\mu\right)\right].
   \label{propagator}
\end{aligned}
\end{equation}

If one sets $\Omega=0$, then it will return to the standard photon propagator in the theory of Maxwell. From Eq.~\eqref{propagator}, we can obtain the dispersion relation for the photon, which is the pole of the propagator, leading to
\begin{equation}
    k^2-\Omega^2=0,
    \label{dispreli}
\end{equation}
or more conveniently written as
\begin{equation}
    E_\gamma^2 = \vec{k}^2 + \Omega^2.
    \label{disprel}
\end{equation}

It can be immediately seen that in the VSR framework the photon acquires an \textit{inertia}, modifying its standard dispersion relation. Hence, its energy is no longer simply proportional to its momentum, but also depends on the parameter $\Omega$. Physically, this means that the photon exhibits a different propagation mode, where it behaves as if it was, effectively, a massive particle. 

More importantly, the directional terms $(s \cdot k)$ in the propagator contribute only as corrections to the polarization structure and do not alter the dispersion relation. Although directional dependence arises from terms proportional to $(s \cdot k)^{-1}$ or other nonlocal effects in VSR, these contributions are strongly suppressed for high-momentum photons, where $|k| \gg \Omega$. Consequently, the leading-order correction introduces a scalar modification to the photon propagator, ensuring that it remains isotropic and dominant in the energy-momentum relation.

The same alteration appears for the fermion propagator \cite{AlfaroSoto}, given by
\begin{equation}
    S_F(p) = i\frac{\not{p} + m_e - \frac{\Omega^2}{2} \left(\slashed{s}/(s \cdot p)\right)}{p^2 - M^2},
    \label{ferprop}
\end{equation}
where $M^2 \equiv m_e^2 + \Omega^2.$ The dispersion relation for the fermion is obtained from the poles of the propagator in Eq.~\eqref{ferprop}, leading to
\begin{equation}
    p^2-M^2=0,
\end{equation}
or,
\begin{equation}
    E^2 = \vec{p}^2 + \underbrace{m_e^2 + \Omega^2}_{M^2}.
\end{equation}

Instead of following the standard dispersion relation, the fermionic particle also acquires an inertia, as if the VSR vacuum behaves as medium, altering the particle's dynamics, by making it heavier. This corroborates with the intriguing feature of a Lorentz violating vacuum behaving similarly to a nontrivial medium \cite{Colladay1998}.

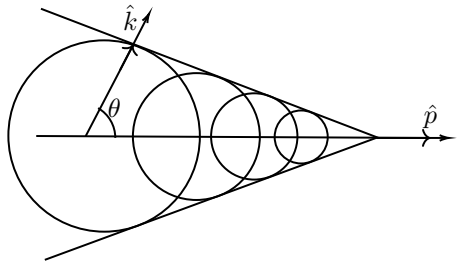
\begin{figure}[h!]
\begin{center}
\tikzset{every picture/.style={line width=0.75pt}} 

\begin{tikzpicture}[x=0.65pt,y=0.65pt,yscale=-0.5,xscale=0.5]

\draw[->] (101,151.5) -- (560,153.75) node[pos=1.0, above] {$\hat{p}$};

\draw[->] (158.5,151.75) -- (214,45.25) node[pos=1.3, left] {$\hat{k}$};

\draw   (68.57,151.36) .. controls (68.46,89.83) and (118.26,40) .. (179.81,40.07) .. controls (241.35,40.15) and (291.32,90.1) .. (291.43,151.64) .. controls (291.54,213.17) and (241.74,263) .. (180.19,262.93) .. controls (118.65,262.85) and (68.68,212.9) .. (68.57,151.36) -- cycle ;

\draw    (106.5,3.25) -- (499.5,153.75);

\draw    (111,295.75) -- (496.5,153.75);

\draw   (213.5,152.13) .. controls (213.5,111.32) and (246.57,78.25) .. (287.38,78.25) .. controls (328.18,78.25) and (361.25,111.32) .. (361.25,152.13) .. controls (361.25,192.93) and (328.18,226) .. (287.38,226) .. controls (246.57,226) and (213.5,192.93) .. (213.5,152.13) -- cycle ;

\draw   (304.5,152) .. controls (304.5,124.25) and (327,101.75) .. (354.75,101.75) .. controls (382.5,101.75) and (405,124.25) .. (405,152) .. controls (405,179.75) and (382.5,202.25) .. (354.75,202.25) .. controls (327,202.25) and (304.5,179.75) .. (304.5,152) -- cycle ;

\draw   (378.5,152.63) .. controls (378.5,135.57) and (392.32,121.75) .. (409.38,121.75) .. controls (426.43,121.75) and (440.25,135.57) .. (440.25,152.63) .. controls (440.25,169.68) and (426.43,183.5) .. (409.38,183.5) .. controls (392.32,183.5) and (378.5,169.68) .. (378.5,152.63) -- cycle ;

\draw    (198.5,74.75) -- (231.58,10.53);
\draw [shift={(232.5,8.75)}, rotate = 117.26] [color={rgb, 255:red, 0; green, 0; blue, 0 }  ][line width=0.75]    (10.93,-3.29) .. controls (6.95,-1.4) and (3.31,-0.3) .. (0,0) .. controls (3.31,0.3) and (6.95,1.4) .. (10.93,3.29);

\draw    (496.5,153.75) -- (579.5,153.75);
\draw [shift={(581.5,153.75)}, rotate = 180] [color={rgb, 255:red, 0; green, 0; blue, 0 }  ][line width=0.75]    (10.93,-3.29) .. controls (6.95,-1.4) and (3.31,-0.3) .. (0,0) .. controls (3.31,0.3) and (6.95,1.4) .. (10.93,3.29);

\draw (173,119) arc[start angle=300, end angle=360, radius=40] node[pos=6, right] {$\theta$};
\end{tikzpicture}
\end{center}
\begin{minipage}{0.9\linewidth}
    \caption{ \justifying A pictorial representation of the Cherenkov cone for a charged particle with momentum $\hat{p}$ emitting a radiation with wavefront vector $\hat{k}$.}
   \label{Cherenkov cone} 
\end{minipage}
\end{figure}

An interesting effect that emerges in macroscopic media is the emission of Cherenkov light \cite{Landau,Masudele}. This phenomenon occurs when a charged particle travels through a medium at a speed greater than the phase velocity of light in that medium. Cherenkov was the first to explore this type of radiation systematically, in 1934 \cite{Cherenkov}. There are two points of view in order to understand the phenomena. From the \textit{microscopic} standpoint, when a charged particle, such as an electron, travels through a medium, it disrupts the electric field of the atoms or molecules in its path. This disruption causes the atoms or molecules to become temporarily polarized. As the polarized particles return to their normal state, they emit electromagnetic radiation in the form of photons. The emitted photons collectively form a cone of light, as in Figure \ref{Cherenkov cone}, that spreads outwards from the path of the charged particle. 

From the \textit{macroscopic} point of view, it is convenient to regard the moving charge as the source of radiation. In this viewpoint, one would have an electron moving in a medium with velocity $u$, and suddenly it emits a photon. While such a process is forbidden in standard Quantum Electrodynamics in the vacuum, it is permitted within a medium, enabling the emission of this radiation.  The derivation of the Cherenkov angle, based on energy and momentum conservation in this scenario, was previously considered in \cite{Cox}, yielding the result:
\begin{equation}
 \cos \theta = \frac{1}{\beta n} + \frac{E_\gamma}{2 |\vec{p}|n}(n^2-1), 
 \label{cosfinal}
\end{equation}
where $\beta = \frac{u}{c},$ $E_\gamma$ is the energy of the photon, $\vec{p}$ is the 3-momentum of the charged particle and $n$ is the refractive index of the medium. It is also important to mention that in the Eq.~\eqref{cosfinal} we have that $c=\hbar=1.$

The first term on the right is the same as the one derived by Tamm and Frank \cite{FT37} in 1937. One can realize that when $n=1$, in other words the vacuum, there is no solution for the Cherenkov effect, for the reason that no particle can travel faster than the speed of \textit{causality}. The last term corresponds to the recoil of the charged particle and, since $E_\gamma$ is proportional to $h$, taking the limit $h \rightarrow{} 0$ would lead us back to the classical regime. As mentioned in \cite{Cox}, the equation Eq.~\eqref{cosfinal} could be rewritten in terms of the wavelength of the electron, resulting in a term proportional $\sim \left(\frac{\Lambda}{\lambda}\right),$ where $\Lambda$ is the wavelength of the electron. Since the wavelength of the electron is much smaller than the photon, this last term is already small, implying that the classical approximation is valid. 

In Lorentz violating theories and non-linear Electrodynamics, there are also proposals for the occurrence of Vacuum Cherenkov radiation  \cite{Macleod,Lehnert,LPRL,Helayel,Altvacuum,AltVacprl,ALTSCHULvac}. However, in these cases, the symmetry structure of the Electrodynamics used is the same as that of Maxwell's theory, i.e., special relativity. As previously asserted, Lorentz violating vacuum can behave as a medium, enabling the Cherenkov effect in what might be described as a ``non-trivial vacuum'', and certainly the deviation from the vacuum refractive index is small. 

As in the standard electrodynamics, the same construction presented in \cite{Cox} can be applied. Although VSR is only invariant under a subgroup of the Lorentz group, it remains invariant under spacetime translations, which ensures that energy and momentum are conserved within this framework.

Therefore, we will examine the case where the particle moves parallel to the preferred direction, in this case $\hat{z}$. Hence, we can write its four-momentum as
\begin{equation}
    p = \left(E_\text{in} , 0, 0, \left| \Vec{p} \right|   \right).
\end{equation}

After emitting a photon as illustrated in Figure \ref{photonemit}, the four-momentum of the system can be written, for the charged particle as:
\begin{equation}
    q = \left(E_\text{out}, -\left| \Vec{q} \right| \sin\phi,0, \left| \Vec{q} \right| \cos \phi \right),
\end{equation} 
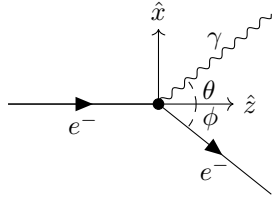
\begin{figure}[h]
  \centering
  \begin{tikzpicture}
  \begin{feynman}
    \draw[->] (0,0) -- (3,0) node[right] {$\hat{z}$};

    \draw[->] (2,0) -- (2,1) node[above] {$\hat{x}$};
    
    \draw[fermion] (0,0) -- (2,0) node[midway,below] {\(e^-\)};
    
    \draw[photon] (2,0) -- ++(1.5,1.2) node[midway,above] {\(\gamma\)};
    
    \draw[fermion] (2,0) -- ++(1.5,-1.2) node[midway,below] {\(e^-\)};
    
    \filldraw (2,0) circle (2pt);
    
    \draw[dashed] (2,0) ++(0.5,0) arc (0:45:0.5) node[midway,right] {\(\theta\)};
    \draw[dashed] (2,0) ++(0.5,0) arc (0:-45:0.5) node[midway,right] {\(\phi\)};
    \end{feynman}
  \end{tikzpicture}
 \begin{minipage}{0.9\linewidth}
    \caption{ \justifying A charged particle, such as an electron, emitting a photon in an angle $\theta$ and being scattered in an angle $\phi$.}
    \label{photonemit}
    \end{minipage}
\end{figure}
and for the photon as:
\begin{equation}
    k = \left(E_\gamma, |\Vec{k}| \sin \theta, 0,|\Vec{k}| \cos \theta \right),
\end{equation}
where we used the natural units $(c=1)$. Furthermore, we have from the conservation of the four-momentum that $p=q+k$. Then,
\begin{equation}
 \left(p-k\right)^2={q}^2, 
 \end{equation}
and if we use that $ E_\text{in}^2 - \left| \Vec{p} \right|^2 = E_{\text{out}}^2 - \left| \Vec{q} \right|^2 = M^2,$ where $M^2 = m_e^2 + \Omega^2,$ we  obtain that
\begin{equation}
0 = E_\gamma^2 -|\vec{k}|^2 - 2p \cdot k,
\label{pck}
\end{equation}
and this equation is similar to the one obtained in \cite{Cox}. 

In a medium where the photon dispersion relation changes to $E=c^\prime|\vec{k}|,$ where $c^\prime=c/n$, with $n$ being the refractive index of the medium, the change implies that the speed of light is reduced to $c^\prime$. For a massive particle, as it is the case for the photon now, this effective mass is intrinsic and does not depend on the medium, hence, this relation is adjusted as follows:
\begin{equation}
  E_\gamma^2 = \left( \frac{|\vec{k}|}{n} \right)^2  + \Omega^2. 
  \label{disprelm}
\end{equation}

Substituting the dispersion relation Eq.~\eqref{disprelm} in Eq.~\eqref{pck}, in order to rewrite $|\vec{k}|$ in terms of the energy of the photon and the parameter $\Omega$, leads to
\begin{equation}
\begin{aligned}
    0=& E_\gamma^2(1-n^2) + n^2 \Omega^2 - 2 E_\text{in}E_\gamma + 2 |\vec{p}|n \sqrt{E_\gamma^2 - \Omega^2}\cos \theta.
    \label{cos1}
\end{aligned}
\end{equation}

The solution to the equation Eq.~\eqref{cos1} for $\cos \theta$ is
\begin{equation}
    \begin{aligned}
        \cos \theta = \frac{E_\gamma(n^2-1)}{2 |\vec{p}|n \sqrt{1 - \left(\frac{\Omega}{E_\gamma} \right)^2}} - \frac{n \Omega^2}{2 |\vec{p}| E_\gamma \sqrt{1 - \left(\frac{\Omega}{E_\gamma} \right)^2}} + \frac{1}{\beta n \sqrt{1 - \left(\frac{\Omega}{E_\gamma} \right)^2}},
    \end{aligned}
\end{equation}
using the fact that $\beta = \frac{|\vec{p}|}{E_\text{in}}.$

Since $\Omega^2$ is a very small parameter, compared to the energy of the photon $E_\gamma$, we can expand the square root as
\begin{equation}
    \begin{aligned}
        \cos \theta = \left(1+ \frac{1}{2}\left(\frac{\Omega}{E_\gamma}\right)^2 \right)\left(\frac{E_\gamma(n^2-1)}{2 |\vec{p}|n} - \frac{n \Omega^2}{2 |\vec{p}| E_\gamma} + \frac{1}{\beta n} \right).
        \label{cosf}
    \end{aligned}
\end{equation}

This equation presents how the Cherenkov angle is altered in the Very Special Relativity scenario. The term $\left(1+\frac{1}{2}\left(\frac{\Omega}{E_\gamma}\right)^2\right)$ in Eq.~\eqref{cosf} emerges from the vacuum of this theory, a property that will be explored in future works. It is straightforward to see that when setting the parameter $\Omega = 0,$ we will return to the standard Eq.~\eqref{cosfinal} derived in \cite{Cox} and, as expected, the deviation of the value of the angle will be very small, since it is proportional to $\Omega^2.$ 

In order to estimate values for this parameter, we will compare it to the Cherenkov angle in the Lorentz invariant framework. The data used to do so is from Particle Identification experiments from the LHCb, specifically the Ring Imaging Cherenkov (RICH) \cite{Lippmann}. The experiment is separated in RICH-1 and RICH-2, that are designed for particle identification across a wide momentum range. RICH-1, located near the interaction point, identifies low-momentum particles around $(2-60)$GeV/c using aerogel and $C_4F_{10}$ gas as radiators. On the other hand, RICH-2 is positioned downstream of the dipole magnet, targeting high-momentum particles $(15-100)$GeV/c using $CF_4$ gas. Both detectors are crucial for the identification of electrons, pions, kaons, protons and muons \cite{Wotton}.

The RICH-1 system employs two different media with distinct refractive indices: aerogel with $n = 1.03$ and $C_4F_{10}$ gas with $n = 1.0014$. The precision of the Cherenkov angle measurement, referred to as the angular resolution ($\sigma_{\theta_C}$), varies depending on the medium. For aerogel, the angular resolution is $\sigma_{\theta_C} = 2.5 \, \text{mrad}$, while for $C_4F_{10}$ gas, it is $\sigma_{\theta_C} = 1.57 \, \text{mrad}$. In comparison, the RICH-2 system achieves higher precision due to its improved angular resolution of $\sigma_{\theta_C} = 0.67 \, \text{mrad}$ and it operates with a refractive index of $n = 1.0005$. A summary of the refractive indices and angular resolutions for the RICH-1 and RICH-2 systems is provided in Table~\ref{table1}.
\begin{table}
    \centering
    \begin{tabular}{ccc}
        \toprule
        \textrm{Refractive index} & 
        \textrm{Max. angle (mrad)} & 
        $\sigma_{\theta_C}$ (mrad) \\
        \midrule
        1.03 & 242 & 2.5 \\
        1.0014 & 53 & 1.57  \\
        1.0005 & 32 & 0.67  \\
        \bottomrule
    \end{tabular}
 \begin{minipage}{0.9\linewidth}
    \caption{ \justifying  This table presents a summary of the variables that we will use from the RICH system. It shows the corresponding maximum Cherenkov angle (Max. angle) and the angular resolution ($\sigma_{\theta_C}$), i.e., the precision of the measurements, for each medium and refractive index.}
  \label{table1}
    \end{minipage}  
\end{table}

The parameter $\Omega$ must be within the range of experimental uncertainty, as any deviation beyond the experiment's precision would have already been detected. Therefore, we can mathematically write this condition as
\begin{equation}
   \Delta \theta \equiv |\theta_{VSR} -\theta_{cl}| \leq \sigma_{{\theta_C}},
    \label{Dcos}
\end{equation}
where $\theta_{VSR}$ is described in Eq.~\eqref{cosf} and $\theta_{cl}$ is described in Eq.~\eqref{cosfinal}. 

The Eq.~\eqref{Dcos} implies that any deviation of the Cherenkov angle from the VSR framework must be smaller than the experimental limitations. Consequently, this allows us to set an upper bound on $\Omega$, which depends on the medium and the particle under consideration.

\begin{figure}[h!]
    \centering
    \includegraphics[width=0.75\linewidth]{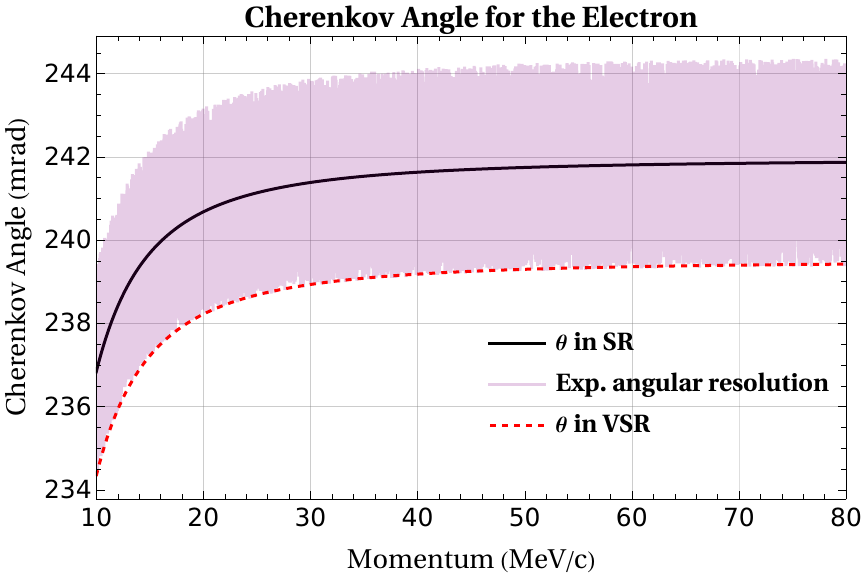}
     \begin{minipage}{0.9\linewidth}
    \caption{ \justifying The Cherenkov angle for the electron vs Momentum, in a medium where $n=1.03$. The purple area is the experiment resolution, generated by a random scan and it is limited by $\sigma_{\theta_C}=2.5 ~\mathrm{mrad}$, which is the experimental limitation of the system. In this graph, $\Omega = 0.207 ~\mathrm{eV/c^2}$. The VSR Cherenkov angle (dashed red line) is at the edge of the detector limitation and the black curve is the standard SR theoretical prediction.}
    \label{limitE}
    \end{minipage}
\end{figure}

For the case of the electron, the SR Cherenkov angle and the VSR modification was plotted for $n=1.03$, in Figure \ref{limitE}. One can observe a more prominent \textbf{decrease} in the Cherenkov angle when we set $\Omega = 0.207 ~\mathrm{eV}/c^2$ i.e. when the VSR Cherenkov angle is just at the border of the experimental limitation, therefore the upper bound obtained for $\Omega \lesssim 0.207 ~\mathrm{eV}/c^2$. For this plot the angular resolution was considered as $\sigma_{\theta_C}=2.5 ~\mathrm{mrad}$.

\begin{figure}[h!]
    \centering
    \begin{minipage}[t]{0.49\linewidth}
        \centering
        \vspace{0pt} 
        \includegraphics[width=0.9\linewidth]{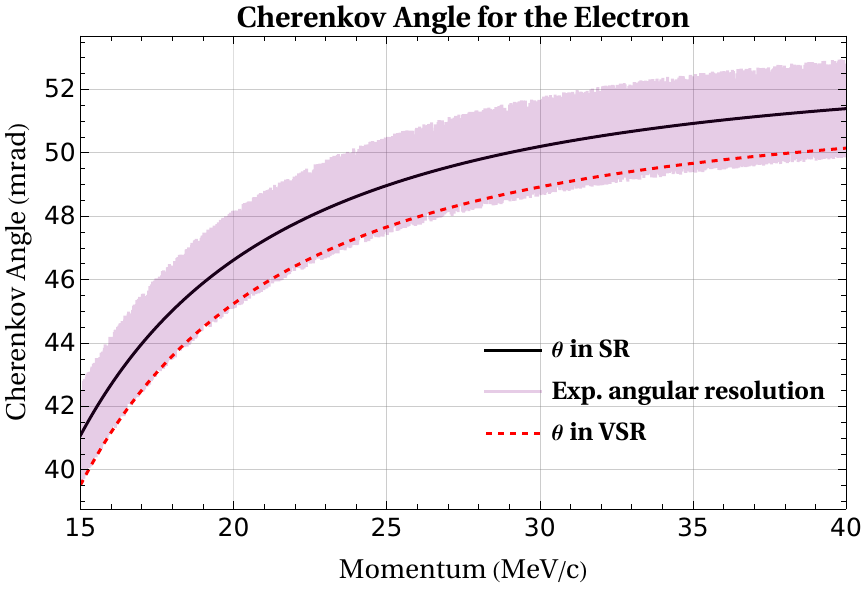}
        \caption{ \justifying Theoretical Cherenkov angle for the electron (thick line) in a medium with $n=1.0014$. The purple area represents the angular resolution region given by $\sigma_{\theta_C} = 1.57 ~\mathrm{mrad}.$ The VSR parameter is limited by $\Omega \lesssim 0.067 ~\mathrm{eV/c^2}$.}
        \label{14}
    \end{minipage}
    \hfill
    \begin{minipage}[t]{0.49\linewidth}
        \centering
        \vspace{0pt} 
        \includegraphics[width=0.9\linewidth]{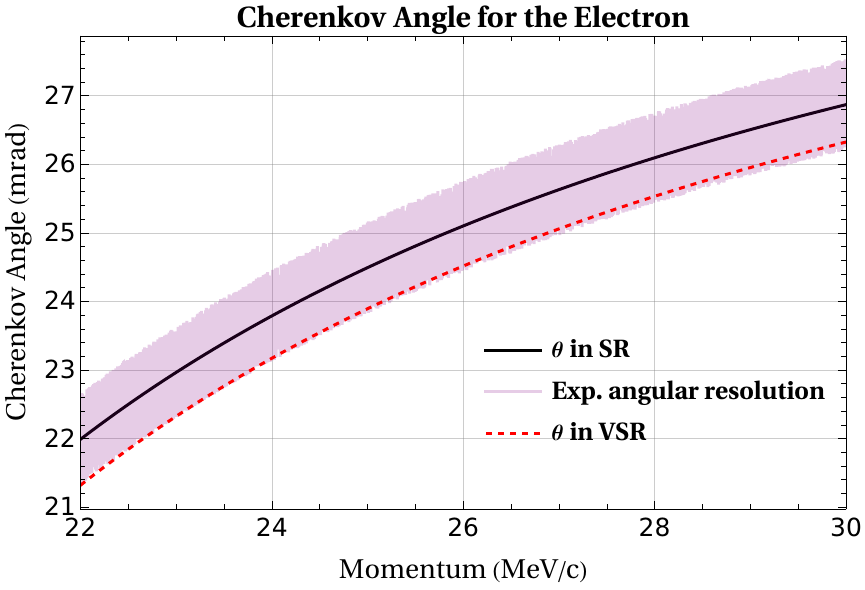}
        \caption{ \justifying Standard theoretical curve (thick line) for the Cherenkov angle in a medium with $n=1.0005$. The purple area represents the experimental uncertainty of $\sigma_{\theta_C} = 0.67 ~\mathrm{mrad}$. The momentum of the particle is measured in $\mathrm{MeV}$, and the VSR parameter was limited to $\Omega \lesssim 0.032 ~\mathrm{eV/c^2}$.}
        \label{05}
    \end{minipage}
\end{figure}

For other refractive indices, tighter bounds on the VSR parameter can be obtained. For instance, in the case of $n = 1.0014$, also for the electron, the upper bound was $\Omega \lesssim 0.067 ~ \mathrm{eV/c^2}$, with an angular resolution of $\sigma_{\theta_C} = 1.57 , \mathrm{mrad}$. This result is illustrated in Figure~\ref{14}. The most stringent limit, for the electron, was achieved when $n = 1.0005$, which had an angular resolution of $\sigma_{\theta_C} = 0.67 ~ \mathrm{mrad}$, significantly more precise than the other media. In this case, the upper bound was $\Omega \lesssim 0.032~\mathrm{eV/c^2}$, as shown in Figure~\ref{05}.

\begin{figure}[h!]
    \centering
    \begin{minipage}[t]{0.49\linewidth}
        \centering
        \vspace{0pt} 
        \includegraphics[width=0.9\linewidth]{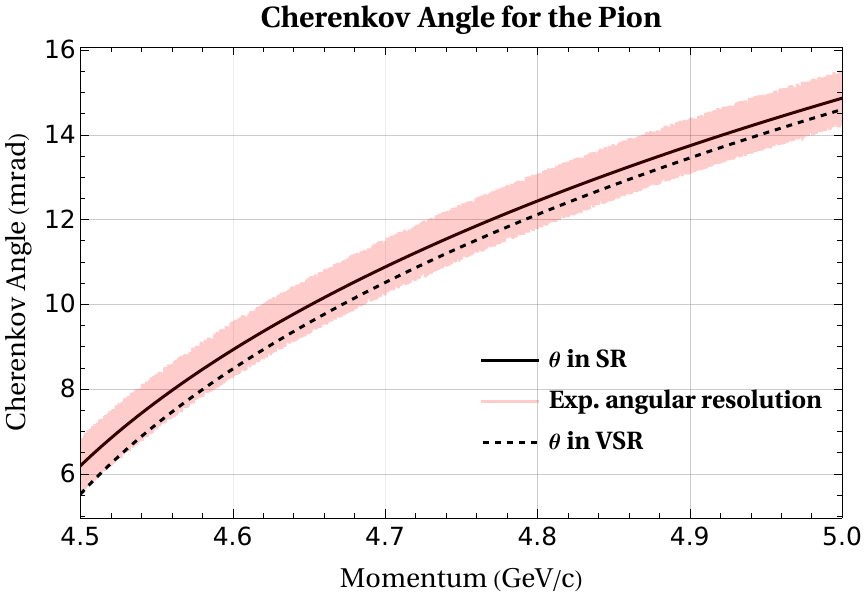}
        \caption{ \justifying Theoretical curve for the Cherenkov angle (thick line) for the pion in a medium with refractive index $n=1.0005$. The red region represents the experiment's uncertainty, where $\Omega$ can assume nonzero values. The lower bound (dashed line) in this case is when $\Omega \lesssim 0.0168 ~\mathrm{eV/c^2}$.}
        \label{pion}
    \end{minipage}
    \hfill
    \begin{minipage}[t]{0.49\linewidth}
        \centering
        \vspace{0pt} 
        \includegraphics[width=0.9\linewidth]{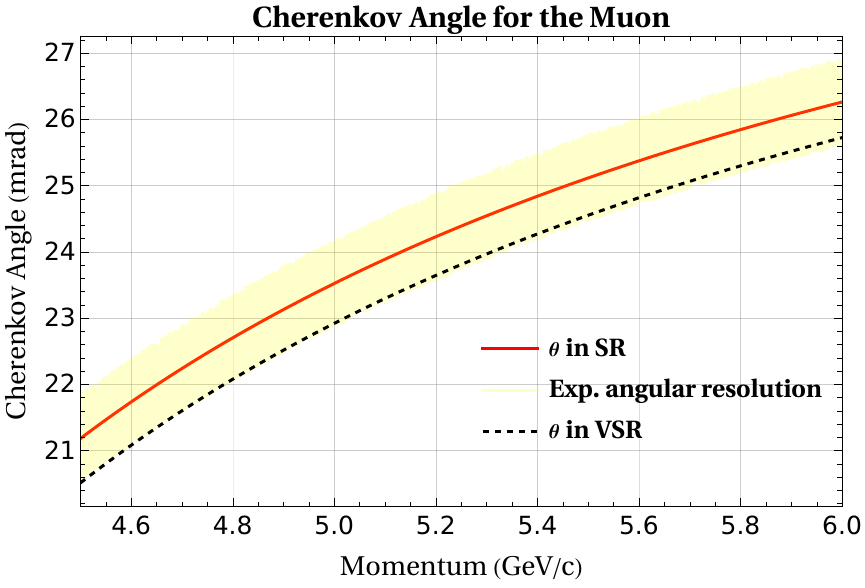}
        \caption{ \justifying Theoretical curve for the Cherenkov angle (red line) for the muon in a medium with refractive index $n=1.0005$. The angular resolution is $\sigma_{\theta_C} = 0.67 ~\mathrm{mrad}$ (yellow area). The lower bound (dashed line) was obtained for $\Omega \lesssim 0.0317 ~\mathrm{eV/c^2}$. The plot illustrates momentum values ranging from $4.5-6 ~\mathrm{GeV}$.}
        \label{muon}
    \end{minipage}
\end{figure}

\begin{figure}[h!]
    \centering
     \begin{minipage}[t]{0.49\linewidth}
        \centering
        \vspace{0pt} 
        \includegraphics[width=0.9\linewidth]{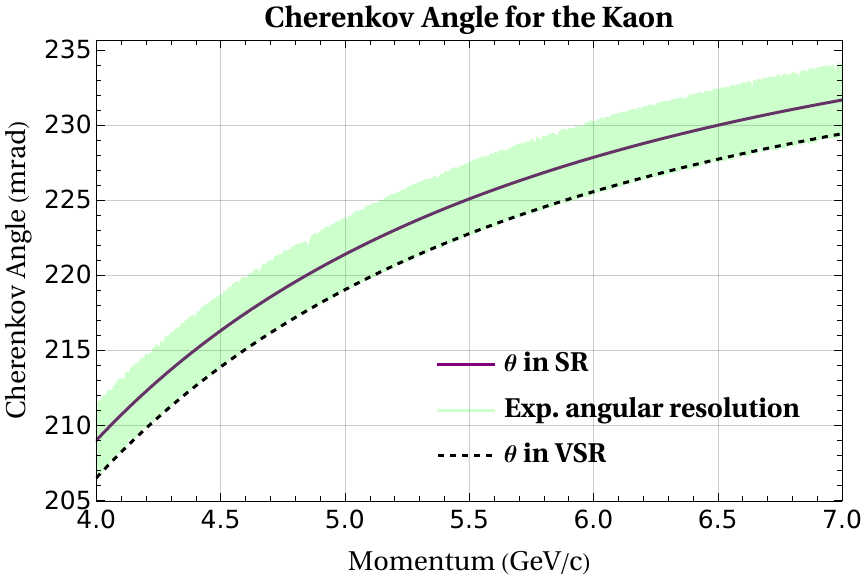}
        \caption{ \justifying Theoretical curve for the Cherenkov angle (purple line) for the kaon in a medium with refractive index $n=1.03$. The lower bound (dashed line) is at $\Omega \lesssim 0.195 ~\mathrm{eV/c^2}$. The experiment’s angular resolution is $\sigma_{\theta_C} = 2.5 ~\mathrm{mrad}$.}
        \label{kaon}
    \end{minipage}
    \hfill
    \begin{minipage}[t]{0.49\linewidth}
        \centering
        \vspace{0pt} 
        \includegraphics[width=0.9\linewidth]{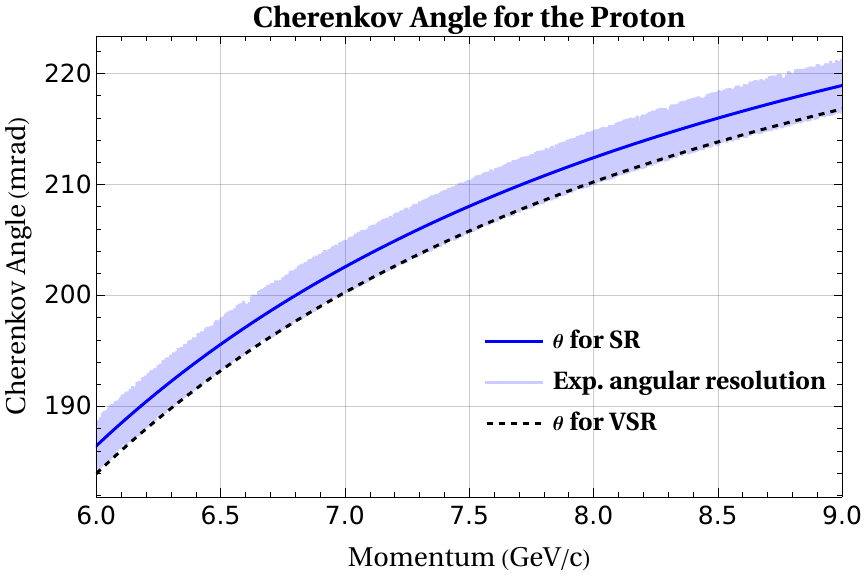}
        \caption{ \justifying Theoretical curve for the Cherenkov angle (blue line) for the proton in a medium with $n=1.03$. The angular resolution is $\sigma_{\theta_C} = 2.5 ~\mathrm{mrad}$, and the VSR Cherenkov angle curve must dwell inside the blue region. The lower bound is when $\Omega \lesssim 0.184 ~\mathrm{eV/c^2}$.}
        \label{proton}
    \end{minipage}
\end{figure}

The same effect can also be observed for other charged particles besides the electron. Figures~\ref{pion}, \ref{muon}, \ref{kaon} and \ref{proton}, and  show plots for the pion, muon, kaon and proton, respectively. The most notable result is for the pion. When $n = 1.0005$ and the angular resolution is $\sigma_{\theta_C} = 0.67 ~ \mathrm{mrad}$, the upper bound on the VSR parameter is $\Omega \lesssim 0.0168 ~ \mathrm{eV/c^2}$, representing the tightest bound among the particles analyzed. We summarize  the results in Table~\ref{table3}, where we organized the calculated upper limits for $\Omega,$ for each media and charged particle.  

\begin{table}[h!]
    \centering
    \begin{tabular}{ccc}
    \hline
        Particle & Refractive index & $\Omega$ bound\\
        \hline
         & 1.03 &  $\lesssim 0.207 ~\mathrm{eV}/c^2$\\
         Electron & 1.0014 &  $ \lesssim 0.067 ~\mathrm{eV/c^2}$\\
          & 1.0005 & $ \lesssim 0.032 ~\mathrm{eV/c^2}$ \\ \hline
          Muon & 1.0005 & $\lesssim 0.0317 ~\mathrm{eV/c^2}$\\ \hline
         Pion & 1.0005 & $ \lesssim 0.0168 ~\mathrm{eV/c^2}$\\ \hline
         Kaon & 1.03 & $ \lesssim 0.195 ~\mathrm{eV/c^2}$ \\ \hline
         Proton & 1.03 & $ \lesssim 0.184 ~\mathrm{eV/c^2}$\\
         \hline
    \end{tabular}
    \begin{minipage}{0.9\linewidth}
    \caption{ \justifying  Summary of the calculated upper bounds on the VSR parameter $\Omega$ for various charged particles across different refractive indices. The tightest constraint, $\Omega \lesssim 0.0168~\mathrm{eV/c^2}$, is achieved for the pion when $n = 1.0005$. These results highlight the dependence of the VSR parameter bound on both the particle mass and the medium’s refractive index. }
  \label{table3}
    \end{minipage} 
\end{table}

To conclude this section, we have shown that within the Very Special Relativity framework, the Cherenkov angle is decreased due to the preferential direction characteristic of VSR, which alters the dispersion relation of particles and their kinematics. Using data from the RICH system of the LHCb experiment, with a refractive index of $n = 1.0005$ and an angular resolution of $\sigma_{\theta_C} = 0.67~\mathrm{mrad}$, we derived an upper limit for the $\Omega$ parameter: $\Omega \lesssim 0.0168~\mathrm{eV}/c^2$. In the next section, we will build upon this result by comparing it with constraints derived from more precise measurements, focusing on the implications of modifications to the speed of light from experiments that test potential deviations of Lorentz invariance.

\section{Confronting Very Special Relativity with Lorentz Invariance Experimental Tests}\label{section 2}

From the dispersion relation of the photon Eq.~\eqref{disprel}, it becomes clear that the speed of light in the VSR framework is altered. A similar equation for the dispersion relation can be found in \cite{Masud,Masud2024} and we will follow the same approach, due to their similarity, in order to obtain an equation for the new speed of light in this scenario.

As the VSR parameter $\Omega$ is independent of the photon momentum $\vec{k}$, and since the momentum values of the experiments considered in this paper ensure that $\vec{k} \gg \Omega$, the group velocity can be derived from the dispersion relation in Eq.~\eqref{disprel} as
\begin{equation}
    v_g = \frac{\partial |E(\vec{k})|}{\partial |\vec{k}|} \approx c \left(1-\frac{1}{2}\frac{\Omega^2}{E_\gamma^2} \right),
\end{equation}
where we have not used natural units for a better interpretation of the equation. Therefore, we obtain a reduction in the speed of light, that depends on the ratio between $\Omega$ and $E_\gamma$. Since, $E_\gamma \gg \Omega$, we anticipate a minimal deviation in $c.$ Additionally, if $\Omega=0,$ we return to the standard speed of light value in the vacuum.

Now, if we use the upper limit for $\Omega \lesssim 0.0168 ~\mathrm{eV/c^2}$ found in the previous section, corresponding to the most restrictive case, the expression for the reduction of the speed of light would indicate a decrease of approximately $0.1\%$, when considering $E_\gamma$ in the order of eV, a deviation that could already have been measured since the most precise measurements of the constancy of the speed of light are at the $(10^{-17})$ level \cite{cred,Lightiso}. Therefore, the constraint on $\Omega$ should be approximately:
\begin{equation}
    \Omega \lesssim 4.47 \times 10^{-9} \times \left(\frac{E_\gamma}{\mathrm{eV}} \right)\frac{\mathrm{eV}}{c^2},
    \label{LIVt}
\end{equation}
which implies that, for instance, if $E_\gamma$ is in electron volts (eV), that $\Omega$ would be constrained to be less than $4.47 \times 10^{-9} ~\mathrm{eV}/c^2$. Consequently, the calculations presented in this work show that the current level of precision in the measurements of the Cherenkov angle is not competitive in probing signatures of Lorentz Invariance Violation within the framework of Very Special Relativity from experiments measuring such effects.

The bound value obtained in Eq.~\eqref{LIVt} is very interesting since, as suggested in \cite{VSRGC}, the $\Omega$ coefficient, as previously asserted, must be of the same order of magnitude as the CP-violating parameter $\bar{\theta} \lesssim 10^{-10}$. This is particularly intriguing because it suggests a convergence of constraints arising from different theoretical and experimental considerations; however, both perspectives point to a similarly small bound value.

Another stringent bound on $\Omega$ arises from certain realizations of VSR \cite{AlfaroSoto}, where the parameter $\Omega$ is connected to the photon mass $m_\gamma$. In this context, $\Omega$ is estimated to be approximately $10^{-16} \, \mathrm{eV}/c^2$ \cite{Phmass}. This bound reinforces the idea that VSR provides a natural setting for exploring phenomena that deviate slightly from traditional Lorentz-invariant theories. 

The summary of the results of this section were organized in Table~\ref{table2}. We can conclude that the upper limits calculated in this section provide significantly tighter constraints on the value of $\Omega$ compared to those derived from Cherenkov effect experiments. This is because RICH detectors are less precise than experiments that measure the constancy of the speed of light, resulting in more flexible bounds on the $\Omega$ value.

In the next section, we will present an analogy between VSR and Minkowski's electrodynamics, where the deviation attributed to VSR will not rely on an abstract parameter $\Omega$, but rather on a concrete physical effect.

\begin{table}
\begin{center}
\begin{tabular}{cc}
    \toprule
    \textrm{Experiments and proposals} & $\Omega$ (eV/$c^2$) \\ 
    \midrule
    \textrm{RICH-1}, $n=1.03$   & $\lesssim 0.183$ \\
    \textrm{RICH-1}, $n=1.0014$ & $\lesssim 0.076$ \\
    \textrm{RICH-2}, $n=1.0005$ & $\lesssim 0.0168$ \\
    Optical cavity & $\lesssim 4.47 \times 10^{-9}$ \\
    $\bar{\theta}$ & $\leq 10^{-10}$ \\
    $m_\gamma$ & $\leq 10^{-16}$ \\
    \bottomrule
\end{tabular}
\end{center}
\begin{minipage}{0.9\linewidth}
    \caption{ \justifying  This table summarizes the upper bounds for the parameter $\Omega$ (in units of eV/$c^2$) based on various experiments and theoretical proposals. The experiments include measurements using RICH detectors with refractive indices $n$ specified for each case, as well as constraints from the photon mass ($m_\gamma$), the $\bar{\theta}$ parameter, and optical cavity experiments. These bounds provide insight into the constraints on $\Omega$ under different experimental setups and theoretical assumptions.}
    \label{table2}
    \end{minipage}
\end{table}

\section{An  analogy with Minkowski's electrodynamics in dielectric media}\label{section 3}

The purpose of this section is to highlight a striking analogy that exists between the formalism analyzed in the foregoing and conventional relativistic electrodynamics in a dielectric medium endowed with a permittivity $\varepsilon$ and a permeability $\mu$ in its rest inertial frame, corresponding to a   refractive index  $n= \sqrt{\varepsilon \mu}$. The  parameters $\varepsilon$ and $\mu$ are assumed to be constants. The Minkowski energy-momentum tensor $T_{\mu\nu}^M$ has the great advantage that it is able to describe all experimental results on radiation pressure on liquids in a straightforward way, and also adjusts itself nicely to a canonical representation. For an introduction to this case of electrodynamics the reader may consult \cite{moller72}. There are also several research papers, for instance \cite{Obukhov,brevik19,brevik20}, and the paper in \cite{brevik17} that analyzes the mentioned canonical representation in terms of a mapping vacuum-medium procedure.

 We consider the system from the laboratory frame, where the medium moves with  a
 uniform four-velocity $V_\mu= ({\bf V}, V_4) = ({\bf V}, iV_0)$, satisfying $V_\mu V_\mu= -1$  (we use here the Minkowski metric, where $x_4=it$). Thus ${\bf V}^2=V_0^2 -1$. The photon four-momentum is spacelike, $k_\mu= ({\bf k}, ik_0)= ({\bf k}, i\omega)$. When  contracted with $V_\mu$, it yields  $k_\mu V_\mu= ({\bf k\cdot V}-\omega V_0)$.

 The equation of motion becomes
\begin{equation}
k_\mu k_\mu -\kappa (k_\mu V_\mu)^2=0, \quad \kappa \equiv  n^2-1, \label{18}
\end{equation}
from which we derive the dispersion relation
\begin{equation}
\omega = \frac{ \kappa V_0({{\bf k\cdot V)}
 \pm \sqrt{ (1+\kappa V_0^2){\bf k}^2-\kappa ({\bf k\cdot V)}^2} }}
{1+\kappa V_0^2}. \label{19}
\end{equation}
There are thus in general two  real roots,  $\omega_1$ and $\omega_2$. We give the expressions for their sum and product,
\begin{equation}
\omega_1+\omega_2= \frac{2\kappa V_0({\bf k\cdot V})}{1+\kappa V_0^2}, \label{20}
\end{equation}
\begin{equation}
\omega_1\omega_2 = \frac{\kappa ({\bf k\cdot V})^2-{\bf k}^2}{1+\kappa V_0^2}. \label{21}
\end{equation}
If  ${\bf (k\cdot V)}>0$ then $\omega_1+\omega_2>0$, and for low and moderate values of ${|\bf V}|, ~\omega_1\omega_2<0$,  so there is one positive and one negative root with a positive sum. If the velocity is high, $\kappa ({\bf k\cdot V})^2 > {\bf k}^2$, then $\omega_1\omega_2 >0$, and there are two positive roots.

Of main interest in the present case is when the velocity is large. Let us assume that ${\bf k}$ has the same direction as ${\bf V}$.  When assuming $V_0^2 \rightarrow \infty, |{\bf V}|^2 \rightarrow \infty$, but still recognizing  the general relation ${\bf V}^2=V_0^2 -1$, we obtain
\begin{equation}
\omega = k\left( 1 \pm \frac{1}{\sqrt{\kappa} V_0^2} \right).
\end{equation}
It shows that there occurs a very small deviation from the relation $\omega = k$. This can be compared to the theory in the previous sections, where the deviation from $\omega = k$ was caused by the introduced parameter $\Omega$; cf. Eq.~\eqref{disprel}. It may be considered as an advantage that the inertia of the photon is in our alternative description  caused by a clear physical effect, the velocity $V_\mu$ of the medium, instead of being attributed to a fictitious parameter   only.

The following characteristics of the Minkowski electrodynamics ought to be noted:

\noindent 1. The Minkowski energy-momentum tensor $T_{\mu\nu}^M$ is not intended to describe a total physical system. Instead, it describes an open (often called a nonclosed) system, consisting of the electromagnetic field in the medium plus its interaction with the matter, but not the matter itself. For that reason, one would expect that the four-divergence of the energy-momentum tensor should be zero. However, the zero condition
\begin{equation}
\partial_\nu T_{\mu\nu}^M =0 \label{divergence}
\end{equation}
holds nevertheless, for all values of $V_\mu$. The implication of this is that for a radiation field, the total energy and momentum constitute a four-vector, implying in turn that the photon four-momentum $k_\mu$ is a four-vector (cf. the discussion on this point in Ref.~\cite{moller72}). Moreover, it turns out that the photon four-momentum is a spacelike vector.

\noindent 2. The Minkowski tensor is nonsymmetric. This is the price one must pay for restricting oneself to an open physical system. The nonsymmetry occurs already in the rest inertial frame, since the Minkowski  photon momentum density is ${\bf D\times B}$
instead of ${\bf E\times H}$, as one would normally expect according to Planck's principle of inertia of energy saying that the momentum density is equal to the Poynting vector divided by $c^2$.  In the rest frame, the nonsymmetry occurs in the fourth line and column in the Minkowski tensor.

In conclusion, we think the analogy with the VSR theory is striking, as both the vanishing of the four-divergence of the Minkowski tensor (conservation of total energy and momentum) and nonconservation of the  angular momentum are properties just encountered  in the VSR theory. Some caution is here needed, however, in not drawing  the analogy too far. The Minkowski theory is one special case only, restricted to  an open physical system as emphasized already, and should not be taken straightaway to encompass the entire Universe.\\

{\Large{\textbf{Special Cases}}}\\

{\it 1.  The case $\kappa = 0 ~(n=1)$:}

Then the dispersion equation (\ref{19}) yields
\begin{equation}
\omega = \pm \sqrt{{\bf k}^2} = \pm {|\bf k}|,
\end{equation}
as it should.

\bigskip
{\it 2. The case of low velocities, $v \ll~1$, but arbitrary $n$:}

Then $V_\mu = \gamma({\bf v}, i) \rightarrow ({\bf v}, i)$,  and one can calculate
\begin{equation}
\kappa V_0({\bf k\cdot V}) =
\kappa {\bf (k\cdot v)},
\end{equation}
\begin{equation}
(1+\kappa V_0^2) {\bf k}^2 = \left( 1+\frac{\kappa}{1-v^2}\right) {\bf k}^2 = (n^2+\kappa {\bf v}^2){\bf k}^2,
\end{equation}
\begin{equation}
\kappa ({\bf k\cdot V})^2 = \kappa ({\bf k\cdot v})^2,
\end{equation}
to the second order. Then, the square root above becomes
\begin{equation}
\sqrt{ (n^2+\kappa {\bf v}^2){\bf k}^2 -  \kappa ({\bf k\cdot v })^2} = |{\bf k}|n,
\end{equation}
and the dispersion equation reduces in this approximation to
\begin{equation}
\omega = \frac{\kappa}{n^2}({\bf k\cdot v}) \pm \frac{|{\bf k}|}{n}.
\end{equation}
Here, the factor $\kappa/n^2$ appears at first sight disturbing, as one might think it comes into conflict with the Lorentz transformation of the wave vector $k_\mu$ at small velocities.  The formalism is however consistent after all, noting that $\omega$ and $\bf k$ refer not to the rest inertial frame of the medium, but to a frame in which the medium has a uniform velocity $\bf v$. Assume for simplicity that the medium moves in the $x$ direction, parallel to $\bf k$. Then the last formula yields
\begin{equation}
\omega = \left( \frac{1}{n}+\frac{\kappa}{n^2}v\right)k_x.
\end{equation}
Let now $\omega^0$ refer to the medium's rest system. In this system, $k_x^0= n\omega^0$. The wave vector component $k_x$ transforms according to the Lorentz transformation as follows:
\begin{equation}
k_x= (n+v)\omega^0.
\end{equation}
This means,
\begin{equation}
\omega = \left( \frac{1}{n}+\frac{\kappa}{n^2}v\right) (n+v)\omega^0 =(1+nv)\omega^0,
\end{equation}
which is in complete agreement with the Lorentz transformation of the frequency. Thus, the scheme is consistent. Also there is no restriction on the magnitude of $n$.

\section*{Conclusions}

In conclusion, we have analyzed the implications of Very Special Relativity (VSR) on the Cherenkov effect and its connection to Lorentz invariance violation. Our findings demonstrate that the Cherenkov angle is reduced within the VSR framework. Specifically, we have derived an upper limit for the $\Omega$ parameter as $\Omega \leq  0.0168~\mathrm{eV}/c^2$, for the pion moving in a medium with a refractive index of $n=1.0005$. This constraint was calculated using data from RICH-2 but appears inconsistent when compared with results from more precise experiments. 

By examining the photon dispersion relation in this context, we found that the speed of light is modified in VSR, resulting in a group velocity dependent on the ratio $\Omega^2/E_\gamma^2$ . Nevertheless, since $\Omega$ has to be extremely small, approximately $\Omega \lesssim 4.47 \times 10^{-9} ~\mathrm{eV}/c^2$, these modifications are negligible and remain undetectable with current experimental precision. The value obtained has many orders of magnitude lower than the value estimated by the Cherenkov angle experiments, due to the enormous precision of measurements on the constancy of the speed of light, therefore, the former offers tighter bounds to the parameter $\Omega$.

Additionally, by correlating $\Omega$ with the photon mass $m_\gamma$ and the CP-violating parameter $\bar{\theta}$, we derived even stricter theoretical constraints on $\Omega$, which are summarized in Table \ref{table2}. These limits indicate, once again, that the effect of $\Omega$ on the speed of light is exceptionally small, further emphasizing the experimental challenges in detecting deviations from Lorentz invariance.

Overall, our results highlight that while VSR introduces interesting theoretical modifications to well established physical principles, the experimental validation of these effects can not be detected nowadays, due to its minuscule values. However, future advances in precision measurement techniques may provide a manner to probe such subtle deviations. 

Lastly, we have highlighted a striking analogy between Very Special Relativity (VSR) and Minkowski's electrodynamics in a dielectric medium. While the analogy has its limitations and cannot be extended indiscriminately, we have demonstrated that the dispersion relation derived from Minkowski's electrodynamics within a dielectric medium is mathematically identical to the VSR dispersion relation for the photon. Importantly, Minkowski's framework offers a more tangible physical interpretation, where this deviation depends on the velocity of the medium $V_\mu,$ and such an analogy could be more explored in future works.

\section*{Acknowledgments}

We are much grateful to Felix Karbstein, Frans Klinkhamer, Alan Kostelecky and  Ralf Lehnert  for several useful discussions. Our special thanks   go  to Brett Altschul, Yuri Obukhov, Per Osland and Anca Tureanu for many  illuminating  discussions and suggestions. B. A. Couto e Silva gratefully acknowledges financial support from FAPEMIG and CAPES.
Appreciation is also extended to the University of Helsinki, particularly the Helsinki Institute of Physics,
for their warm hospitality. B. L. Sánchez-Vega thanks the National Council for Scientific and
Technological Development of Brazil, CNPq, for the financial support through grant n◦ 311699/2020-0.

\bibliographystyle{utphys} 
\bibliography{apssamp.bib} 

\providecommand{\href}[2]{#2}\begingroup\raggedright\begin{thebibliography}{10}

\bibitem{ProceedingsL}
R.~Lehnert, {\em Proceedings of the Ninth Meeting on CPT and Lorentz Symmetry}.
\newblock World Scientific Publishing Co. Pte. Ltd., 2023.

\bibitem{LeeYang1956}
T.~D. Lee and C.~N. Yang, ``Question of parity conservation in weak interactions,'' \href{https://dx.doi.org/10.1103/PhysRev.104.254}{{\em Physical Review} {\bfseries 104} no.~1, (1956) 254--258}.

\bibitem{Wu1957}
C.~S. Wu, E.~Ambler, R.~W. Hayward, D.~D. Hoppes, and R.~P. Hudson, ``Experimental test of parity conservation in beta decay,'' \href{https://dx.doi.org/10.1103/PhysRev.105.1413}{{\em Physical Review} {\bfseries 105} no.~4, (1957) 1413--1415}.

\bibitem{Seiberg1999}
N.~Seiberg and E.~Witten, ``String theory and noncommutative geometry,'' \href{https://dx.doi.org/10.1088/1126-6708/1999/09/032}{{\em Journal of High Energy Physics} {\bfseries 1999} no.~09, (Oct, 1999) 032}. \url{https://dx.doi.org/10.1088/1126-6708/1999/09/032}.

\bibitem{klink}
F.~R. Klinkhamer and M.~Schreck, ``New two-sided bound on the isotropic lorentz-violating parameter of modified maxwell theory,'' \href{https://dx.doi.org/10.1103/PhysRevD.78.085026}{{\em Phys. Rev. D} {\bfseries 78} (2008) }. \url{https://link.aps.org/doi/10.1103/PhysRevD.78.085026}.

\bibitem{Coleman1999}
S.~Coleman and S.~L. Glashow, ``High-energy tests of {L}orentz invariance,'' {\em Phys. Rev. D} {\bfseries 59} (1999) 116008.

\bibitem{VSRGC}
A.~G. Cohen and S.~L. Glashow, ``Very special relativity,'' \href{https://dx.doi.org/10.1103/PhysRevLett.97.021601}{{\em Phys. Rev. Lett.} {\bfseries 97} (2006) }.

\bibitem{Tureanu}
M.~M. Sheikh-Jabbari and A.~Tureanu, ``Realization of {C}ohen-{G}lashow {V}ery {S}pecial {R}elativity on {N}oncommutative {S}pace-{T}ime,'' \href{https://dx.doi.org/10.1103/PhysRevLett.101.261601}{{\em Phys. Rev. Lett.} {\bfseries 101} (Dec, 2008) 261601}. \url{https://link.aps.org/doi/10.1103/PhysRevLett.101.261601}.

\bibitem{CPPRL}
{\bfseries BABAR} Collaboration, B.~Aubert, D.~Boutigny, {\em et~al.}, ``Measurement of $\mathit{CP}$-violating {A}symmetries in ${B}^{0}$ decays to $\mathit{CP}$ {E}igenstates,'' \href{https://dx.doi.org/10.1103/PhysRevLett.86.2515}{{\em Phys. Rev. Lett.} {\bfseries 86} (2001) }.

\bibitem{CPPRL2}
{\bfseries LHCb} Collaboration, R.~Aaij, C.~Abell\'an~Beteta, {\em et~al.}, ``Observation of {$CP$} {V}iolation in {C}harm {D}ecays,'' \href{https://dx.doi.org/10.1103/PhysRevLett.122.211803}{{\em Phys. Rev. Lett.} {\bfseries 122} (2019) }.

\bibitem{CGnm}
A.~G. Cohen and S.~L. Glashow, ``{A Lorentz-Violating Origin of Neutrino Mass?},'' {\em arXiv} (2006) , \href{https://arxiv.org/abs/hep-ph/0605036}{{\ttfamily hep-ph/0605036}}.

\bibitem{Adamson2011}
P.~Adamson, C.~Andreopoulos, R.~Armstrong, and et~al., ``Measurement of the neutrino mass splitting and flavor mixing by {MINOS},'' \href{https://dx.doi.org/10.1103/PhysRevLett.106.181801}{{\em Physical Review Letters} {\bfseries 106} no.~18, (2011) 181801}.

\bibitem{Takeuchi2012}
Y.~Takeuchi and T.~{S}uper Kamiokande~{C}ollaboration, ``Results from {S}uper-{K}amiokande,'' \href{https://dx.doi.org/10.1016/j.nuclphysbps.2012.09.013}{{\em Nuclear Physics B} {\bfseries 229-232} (2012) 79--84}.

\bibitem{Gouvea}
A.~Gouvea, ``Neutrino mass models,'' \href{https://dx.doi.org/10.1146/ANNUREV-NUCL-102115-044600}{{\em Annual Review of Nuclear and Particle Science} {\bfseries 66} (2016) 197--217}.

\bibitem{repspin}
J.~Fan, W.~D. Goldberger, and W.~Skiba, ``Spin dependent masses and {S}im(2) symmetry,'' \href{https://dx.doi.org/https://doi.org/10.1016/j.physletb.2007.03.055}{{\em Physics Letters B} {\bfseries 649} no.~2, (2007) 186--190}. \url{https://www.sciencedirect.com/science/article/pii/S0370269307004194}.

\bibitem{Cox}
R.~T. Cox, ``Momentum and energy of photon and electron in the \ifmmode \check{C}\else \v{C}\fi{}erenkov radiation,'' {\em Phys. Rev.} {\bfseries 66} (1944) 106--107.

\bibitem{Cheon2009}
S.~Cheon, C.~Lee, and S.~J. Lee, ``Sim(2)-invariant modifications of electrodynamic theory,'' {\em Physics Letters B} {\bfseries 679} no.~1, (2009) 73--76.

\bibitem{Bauer2001}
C.~W. Bauer, S.~Fleming, D.~Pirjol, and I.~W. Stewart, ``An effective field theory for collinear and soft gluons: Heavy to light decays,'' \href{https://dx.doi.org/10.1103/PhysRevD.63.114020}{{\em Phys. Rev. D} {\bfseries 63} (May, 2001) 114020}. \url{https://link.aps.org/doi/10.1103/PhysRevD.63.114020}.

\bibitem{AlfaroSoto}
J.~Alfaro and A.~Soto, ``Photon mass in very special relativity,'' \href{https://dx.doi.org/10.1103/PhysRevD.100.055029}{{\em Phys. Rev. D} {\bfseries 100} (2019) 055029}.

\bibitem{Colladay1998}
D.~Colladay and V.~A. Kostelecký, ``Lorentz-violating extension of the standard model,'' {\em Phys. Rev. D} {\bfseries 58} (1998) 116002.

\bibitem{Landau}
L.~D. Landau and E.~M. Lifshitz, \href{https://dx.doi.org/10.1016/B978-0-08-030275-1.50007-2}{{\em Electrodynamics of Continuous Media}}.
\newblock Pergamon Press, England, 2nd~ed., 1984.
\newblock See Chapter 14.

\bibitem{Masudele}
M.~Chaichian, I.~Merches, D.~Radu, and A.~Tureanu, {\em Electrodynamics: An Intensive Course}, ch.~7, p.~420.
\newblock Springer, 2016.

\bibitem{Cherenkov}
P.~Cherenkov, ``{Visible Emission of Clean Liquids by Action of $\gamma$ Radiation},'' {\em Doklady Akad. Nauk SSSR 2} {\bfseries 451} (1934) .

\bibitem{FT37}
I.~Tamm, I.~Frank, ``{Coherent visible radiation from fast electrons passing through matter},'' {\em C. R. Acad. Sci. USSR} {\bfseries 14} (1937) 109--114.

\bibitem{Macleod}
A.~J. Macleod, A.~Noble, and D.~A. Jaroszynski, ``Cherenkov radiation from the quantum vacuum,'' \href{https://dx.doi.org/10.1103/PhysRevLett.122.161601}{{\em Phys. Rev. Lett.} {\bfseries 122} (Apr, 2019) 161601}. \url{https://link.aps.org/doi/10.1103/PhysRevLett.122.161601}.

\bibitem{Lehnert}
R.~Lehnert and R.~Potting, ``\ifmmode \check{C}\else \v{C}\fi{}erenkov effect in lorentz-violating vacua,'' {\em Phys. Rev. D} {\bfseries 70} (2004) .

\bibitem{LPRL}
R.~Lehnert and R.~Potting, ``Vacuum \ifmmode \check{C}\else \v{C}\fi{}erenkov radiation,'' {\em Phys. Rev. Lett.} {\bfseries 93} (2004) .

\bibitem{Helayel}
P.~Gaete and J.~A. Helay\"el-Neto, ``{Vacuum material properties and Cherenkov radiation in logarithmic electrodynamics},'' \href{https://dx.doi.org/10.1140/epjc/s10052-023-11280-w}{{\em Eur. Phys. J. C} {\bfseries 83} no.~2, (2023) 128}, \href{https://arxiv.org/abs/2205.03252}{{\ttfamily arXiv:2205.03252 [hep-ph]}}.

\bibitem{Altvacuum}
B.~Altschul, ``\ifmmode \check{C}\else \v{C}\fi{}erenkov radiation in a {L}orentz-violating and birefringent vacuum,'' \href{https://dx.doi.org/10.1103/PhysRevD.75.105003}{{\em Phys. Rev. D} {\bfseries 75} (May, 2007) 105003}. \url{https://link.aps.org/doi/10.1103/PhysRevD.75.105003}.

\bibitem{AltVacprl}
B.~Altschul, ``Vacuum \ifmmode \check{C}\else \v{C}\fi{}erenkov radiation in {L}orentz-violating theories without {CPT} violation,'' \href{https://dx.doi.org/10.1103/PhysRevLett.98.041603}{{\em Phys. Rev. Lett.} {\bfseries 98} (Jan, 2007) 041603}. \url{https://link.aps.org/doi/10.1103/PhysRevLett.98.041603}.

\bibitem{ALTSCHULvac}
B.~Altschul, ``Finite duration and energy effects in {L}orentz-violating vacuum \ifmmode \check{C}\else \v{C}\fi{}erenkov radiation,'' \href{https://dx.doi.org/https://doi.org/10.1016/j.nuclphysb.2007.12.012}{{\em Nuclear Physics B} {\bfseries 796} no.~1, (2008) 262--273}. \url{https://www.sciencedirect.com/science/article/pii/S0550321307009509}.

\bibitem{Lippmann}
C.~Lippmann, ``Particle identification,'' {\em Nuclear Instruments and Methods in Physics Research Section A: Accelerators, Spectrometers, Detectors and Associated Equipment} {\bfseries 666} (2012) 148--172.

\bibitem{Wotton}
S.~Wotton and {LHCb RICH Collaboration}, ``The {LHC}b {RICH} upgrade for the high luminosity {LHC} era,'' \href{https://dx.doi.org/https://doi.org/10.1016/j.nima.2023.168824}{{\em Nuclear Instruments and Methods in Physics Research Section A: Accelerators, Spectrometers, Detectors and Associated Equipment} {\bfseries 1058} (2024) 168824}.

\bibitem{Masud}
I.~Brevik, M.~Chaichian, and M.~Oksanen, ``{Dispersion of light traveling through the interstellar space, induced and intrinsic Lorentz invariance violation},'' \href{https://dx.doi.org/10.1140/epjc/s10052-021-09707-3}{{\em Eur. Phys. J. C} {\bfseries 81} no.~10, (2021) 926}, \href{https://arxiv.org/abs/2101.00954}{{\ttfamily arXiv:2101.00954 [astro-ph.HE]}}.

\bibitem{Masud2024}
I.~H. Brevik, M.~M. Chaichian, and A.~Tureanu, ``Below the {S}chwinger critical magnetic field value, quantum vacuum and gamma-ray bursts delay,'' {\em In-Print} (2025) .

\bibitem{cred}
S.~Herrmann, A.~Senger, K.~M\"ohle, M.~Nagel, E.~V. Kovalchuk, and A.~Peters, ``Rotating optical cavity experiment testing lorentz invariance at the ${10}^{\ensuremath{-}17}$ level,'' \href{https://dx.doi.org/10.1103/PhysRevD.80.105011}{{\em Phys. Rev. D} {\bfseries 80} (Nov, 2009) 105011}.

\bibitem{Lightiso}
C.~Eisele, A.~Y. Nevsky, and S.~Schiller, ``Laboratory test of the isotropy of light propagation at the ${10}^{\ensuremath{-}17}$ level,'' \href{https://dx.doi.org/10.1103/PhysRevLett.103.090401}{{\em Phys. Rev. Lett.} {\bfseries 103} (Aug, 2009) 090401}. \url{https://link.aps.org/doi/10.1103/PhysRevLett.103.090401}.

\bibitem{Phmass}
L.~Davis, A.~S. Goldhaber, and M.~M. Nieto, ``Limit on the photon mass deduced from pioneer-10 observations of {J}upiter's magnetic field,'' \href{https://dx.doi.org/10.1103/PhysRevLett.35.1402}{{\em Phys. Rev. Lett.} {\bfseries 35} (Nov, 1975) 1402--1405}. \url{https://link.aps.org/doi/10.1103/PhysRevLett.35.1402}.

\bibitem{moller72}
C.~M{\o}ller, {\em The Theory of Relativity}.
\newblock Clarendon Press, Oxford, 2nd~ed., 1972.

\bibitem{Obukhov}
Y.~Obukhov, ``Electromagnetic energy and momentum in moving media,'' \href{https://dx.doi.org/https://doi.org/10.1002/andp.200852009-1012}{{\em Annalen der Physik} {\bfseries 520} no.~9-10, (2008) }. \url{https://onlinelibrary.wiley.com/doi/abs/10.1002/andp.200852009-1012}.

\bibitem{brevik19}
I.~Brevik, ``Spacelike character of the {M}inkowski four-momentum in analog gravity,'' \href{https://dx.doi.org/10.1103/PhysRevA.100.032109}{{\em Physical Review A} {\bfseries 100} no.~3, (2019) 032109}.

\bibitem{brevik20}
I.~Brevik, ``Classical and quantal aspects of {M}inkowski's four-momentum in analog gravity,'' \href{https://dx.doi.org/10.1103/PhysRevA.102.052201}{{\em Physical Review A} {\bfseries 102} no.~5, (2020) 052201}.

\bibitem{brevik17}
I.~Brevik, ``Minkowski momentum resulting from a vacuum–medium mapping procedure, and a brief review of {M}inkowski momentum experiments,'' \href{https://dx.doi.org/https://doi.org/10.1016/j.aop.2017.01.009}{{\em Annals of Physics} {\bfseries 377} (2017) 10--21}.

\end{thebibliography}\endgroup
\end{document}